\documentclass[prl,aps,twocolumn,showpacs,a4paper]{revtex4}
\usepackage{bm}
\usepackage[dvips]{graphicx}
\usepackage{amsmath}
\usepackage{verbatim}
\begin{document}
\newcommand{\la}{\label}
\newcommand{\be}{\begin{equation}}
\newcommand{\ee}{\end{equation}}
\newcommand{\der}{\partial}
\newcommand{\w}{\tilde}
\newcommand{\diag}{{\mathrm{diag}}}
\newcommand{\sgn}{\,{\mathrm{sgn}}}
\newcommand{\str}{{\mathrm{str}}}
\newcommand{\tr}{\,{\mathrm{tr}}\,}
\newcommand{\re}{{\mathrm{Re}}\,}
\newcommand{\im}{{\mathrm{Im}}\,}
\newcommand{\e}{_{\mathrm{eq}}}

\title{Criticality without self-similarity: a 2D system with random long-range hopping}

\author{A. Ossipov$^{1}$, I. Rushkin$^{1}$, E. Cuevas$^2$}
\affiliation{$^1$School of Mathematical Sciences, University of Nottingham, Nottingham NG7 2RD, United Kingdom\\
$^2$Departamento de F\'{\i}sica, Universidad de Murcia, E-30071 Murcia, Spain
}

\date{\today}

\begin{abstract}
We consider a simple model of quantum disorder in two dimensions, characterized by a long-range site-to-site hopping. The system undergoes a metal-insulator transition -- its eigenfunctions change from being extended to being localized. We demonstrate that at the point of the transition the eigenfunctions do not become fractal. Their density moments do not scale as a power of the system size. Instead, in one of the considered limits our result suggests a power of the logarithm of the system size. In this regard, the transition differs from a similar one in the one-dimensional version of the same system, as well as from the conventional Anderson transition in more than two dimensions.
\end{abstract}



\maketitle

Critical wave functions have been a subject of intense theoretical research during the last two decades (see Ref.~\cite{EM08} for a recent review). Recently, they have also attracted a lot of attention from experimentalists \cite{Huse, Lagendijk, Lemarie}. Starting with the innovative work by Wegner \cite{W80}, it has become customary to characterize the critical wave functions by their {\it multifractal} dimensions $d_q$, which determine the scaling of the moments $I_q$ of the wave functions with system size $L$:
\be \label{standard_scaling}
I_q=L^{d}\left\langle|\psi_n({\bf r})|^{2q}\right\rangle\propto L^{-d_q(q-1)}.
\ee
Here the averaging is performed over different disorder realizations, as well as over a small energy window.
The power-law dependence of the moments with exponents $d_q\neq 0$ different from the dimensionality of the space $d$ implies a {\it self-similarity} of fluctuations of the wave function amplitudes on different spatial scales.

Among all quantum disordered systems showing critical behavior, two-dimensional (2D) systems occupy a special place. It is commonly believed that at the critical point such systems possess conformal invariance and should be described by conformal field theories (CFTs), similar to conventional phase transitions in 2D. Although no direct mapping of any disordered quantum system onto a CFT is available at the moment, this conjecture of conformal invariance imposes certain constraints on correlation functions and critical exponents. The validity of these predictions has been checked by numerical simulations for various 2D disordered systems \cite{Obuse, EMM08}, and no significant deviations have been found so far. One such prediction is a relation between the corner and the surface multifractalities \cite{Cardy}:
\be\label{conformal}
\Delta_q^{{\rm c}}(\theta)=\frac{\pi}{\theta}\Delta_q^{{\rm s}},
\ee
where the exponents $\Delta_q^{{\rm x}}=(d_q^{{\rm x}}-d)(q-1)$ are determined by the deviation of the corresponding fractal dimensions $d_q^{\rm x}$ from the dimensionality of space $d$. To define $d_q^{{\rm x}}$ one should take the position of the point ${\bf r}$ in Eq. (\ref{standard_scaling}) to be in a corner of the system boundary with an internal angle $\theta$ (${\rm x}={\rm c}$), on the surface (${\rm x}={\rm s}$), or else in the bulk ( ${\rm x}={\rm b}$).

In the present work we study a random matrix model describing a long-range hopping of a particle on a two-dimensional lattice. Similar to its one-dimensional (1D) counterpart \cite{MFD96}, the model undergoes the Anderson metal-insulator transition. At the transition point, we compute the fractal dimensions perturbatively in the regime of strong criticality and find that
\be
d_q^{{\rm c}}(\theta)=\frac{\theta}{\pi}d_q^{{\rm s}}.
\ee
The dependence on $\theta$ is {\it inverse} to that predicted by CFT in Eq. (\ref{conformal}).
In the opposite regime of weak criticality, the scaling of the moments of the wave functions is found to be consistent not with the standard power-law (\ref{standard_scaling}), but with a {\it logarithmic} dependence
\be\label{ln_scaling}
I_q\propto L^{-2(q-1)}\ln^{\nu_q(q-1)} L.
\ee
Consequently, our model is a rather surprising example of a disordered 2D critical system, whose wave functions are not self-similar and are less localized than the standard critical states.
We believe that the origin of this peculiarity of our model lies in the long-range nature of the Hamiltonian \cite{Cardy}. Therefore, our results might be relevant for other two-dimensional disordered critical systems with a long-range interaction. One important example of such a system is the quantum Hall effect with unscreened Coulomb interactions. Below we present an exact definition of our model, derivations of the main equations and results of the numerical simulations supporting our findings.

\paragraph{The model.} The random matrix model considered here is defined by a random Hamiltonian matrix $H_{{\bf r},{\bf r}'}$, which for simplicity can be taken in the coordinate representation: ${\bf r}$ and ${\bf r}'$ are two points in the system and $H_{{\bf r},{\bf r}'}$ is the amplitude of a quantum hop between them, random due to the disorder in the system. To make the matrix formulation meaningful, it is understood that ${\bf r}$ belongs to a spatial lattice. The Hamiltonian matrix can be either real and symmetric (for a system with time-reversal invariance), or complex and Hermitian (for a system without it). In the standard random matrix theory, all the independent matrix elements $H^{(0)}_{{\bf r},{\bf r}'}$ are taken to be independent Gaussian random variables with zero mean and variance $\epsilon$. This means that the typical hopping amplitude is $\epsilon$ for any two points. One can consider a more elaborate system in which the hopping amplitude decays with the distance:
\be\label{ham}
H_{{\bf r},{\bf r}'}=H^{(0)}_{{\bf r},{\bf r}'}\sqrt{a_{{\bf r},{\bf r}'}}.
\ee
We consider the case when the variance matrix depends only on the distance, i.e. $a_{{\bf r},{\bf r}'}=a(|{\bf r}-{\bf r}'|)$, and normalize the diagonal terms $a(0)=1$. It is well-known that if $a(x)$ decays at infinity faster than any power, all eigenfunctions of $H$ are localized. In the limit of a diagonal matrix $a$, this is a trivial remark. However, we are interested in the situation when the decay is a power-law beyond a certain bandwidth $b$:
\be a(x)=\frac{1}{1+(\frac{x}{b})^{2\alpha}}.\ee
In 1D, this model has been studied in depth \cite{MFD96, EM08}. It is known that the system undergoes a metal-insulator transition at $\alpha=1$: the eigenfunctions are localized for $\alpha>1$ (with power-law tails) and extended for $\alpha<1$, irrespective of $b$. There is strong evidence that the eigenfunctions are multifractal at $\alpha=1$, as in the usual 3D Anderson transition. Being 1D, this model has an obvious advantage for calculations and hence it is widely used as a playground for investigating various features of the Anderson transition \cite{EM08, CK07, KOYC10}.

A similar transition at $\alpha=2$ is believed to exist in 2D \cite{MFD96}, where it is tempting to make a connection between it and a {\it bona fide} Anderson metal-insulator transition in $2+\epsilon$ dimensions. Based on the numerical evidence \cite{PS02}, it is certain that the eigenfunctions in the 2D power-law model are localized for $\alpha>2$ and extended for $\alpha<2$. The question remains: are the eigenfunctions fractal at $\alpha=2$? Since this was the case in 1D, as well as in the true Anderson transition, the same was expected here. Moreover, some numerical experiments pointed in this direction, and even suggested the values of the multifractal dimensions \cite{C04}. However, this numerical work was done only for moderate system sizes and, based on the results, it is not easy to judge whether or not the obtained scaling relation is real. To answer this question we calculate below $I_q$ in the limits of {\it strong} ($b\ll1$) and {\it weak} ($b\gg1$) criticality.

\begin{figure}[t]
\begin{center}
\includegraphics[clip=true,width=\columnwidth]{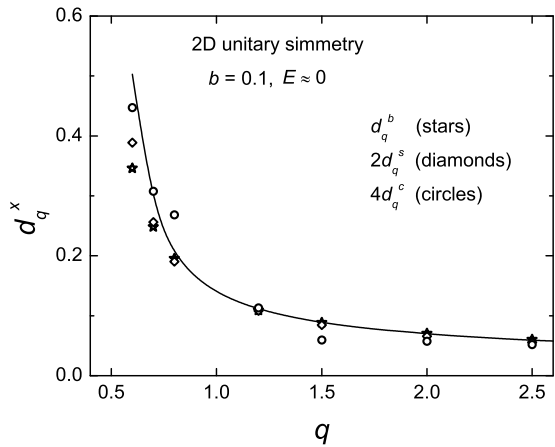}
\end{center}
\caption{Dependence of the corner ($\theta=\frac{\pi}{2}$), surface and bulk fractal dimensions on $q$. Error bars are smaller than the point size. The solid line represents the theoretical prediction (\ref{corner}). }
\label{figure_strong}
\end{figure}

\paragraph{Strong criticality $(b\ll 1)$.}  In this regime the Hamiltonian matrix $H_{{\bf r},{\bf r}'}$ is almost diagonal and we can find $I_q$ by virial expansion. The leading order result for $q>1/2$, valid for any random matrix with independent Gaussian elements, reads \cite{FOR09}
\be
I_q({\bf r})= 1-\eta\frac{\Gamma(q-\frac{1}{2})}{\Gamma(q-1)}\sum_{{\bf r}'}\sqrt{a_{{\bf r},{\bf r}'}} +O(b^4),
\ee
where $\eta=\frac{2}{\sqrt\pi}$ for orthogonal symmetry, $\eta=\sqrt\frac{\pi}{2}$ for unitary symmetry.
Let us assume that the system occupies a  sector of a disk of radius $L$, with the point ${\bf r}$ as its apex and the apex angle $\theta$. We replace the sum with an integral: $\sum_{{\bf r}'}\sqrt{a_{{\bf r},{\bf r}'}}=\theta\int_0^L\Bigl(1+\frac{r^4}{b^4}\Bigr)^{-1/2}rdr=\frac{\theta b^2}{2}\sinh^{-1}\frac{L^2}{b^2}$, and upon sending $L\to\infty$,
\be\label{Iq_strong}
I_q({\bf r})= 1-\eta b^2\frac{\Gamma(q-\frac{1}{2})}{\Gamma(q-1)}\theta \ln L +O(b^4).
\ee
Comparing this result with the expected power-law scaling $I_q\propto L^{-d_q(q-1)}=L(1-(q-1)d_q\ln L+\ldots)$ we can see that the presence of the $\ln L$ term in (\ref{Iq_strong}) is consistent with the power-law (\ref{standard_scaling}), and the fractal dimension is given by
\be\label{corner}
d_q^c(\theta)=\eta b^2\frac{\Gamma(q-\frac{1}{2})}{\Gamma(q)}\theta +O(b^4).
\ee
The expressions for the surface and the bulk fractal dimensions can be obtained by taking $\theta=\pi$ and $\theta=2\pi$, respectively:
\be\label{dq_strong}
d_q^{{\rm c}}(\theta)=\frac{\theta}{\pi}d_q^{{\rm s}}=\frac{\theta}{2\pi}d_q^{{\rm b}}.
\ee
The dependence on $\theta$ is {\it inverse} to the conformal prediction (\ref{conformal}). This result rules out the conformal invariance for our model at the critical point.
In order to check our predictions Eqs. (\ref{corner},\ref{dq_strong}) we carried
out numerical calculations on square samples with lateral
sizes up to $L = 181$ for unitary symmetry. Eigenfunctions of the Hamiltonian matrix
(\ref{ham}) (with the bandwidth $b=0.1$) were obtained by standard diagonalization
subroutines, and the ensemble average was performed over at least
$5\times 10^5$ eigenstates with energies close to zero for each size $L$.

The fractal dimensions $d_q$ were calculated from the scaling
of the moments $I_q$ of the wave functions.
In order to compute them we considered wave
function amplitudes in the following regions of the $L\times L$ sample:
i) $3\times 3$ squares (nine sites) at each corner for $d_q^{{\rm c}}$,
ii) stretches of $L/6$ sites centered in the middle of an edge for $d_q^{{\rm s}}$,
and iii) an $L/6\times L/6$ square in the center
of the sample for $d_q^{{\rm b}}$. The results are
presented in  Fig.~\ref{figure_strong}. The multifractal dimensions $d_q^{{\rm x}}$
are close to the theoretical predictions given by Eqs. (\ref{corner},\ref{dq_strong}).


\begin{figure}[t]
\includegraphics[clip=true, width=\columnwidth]{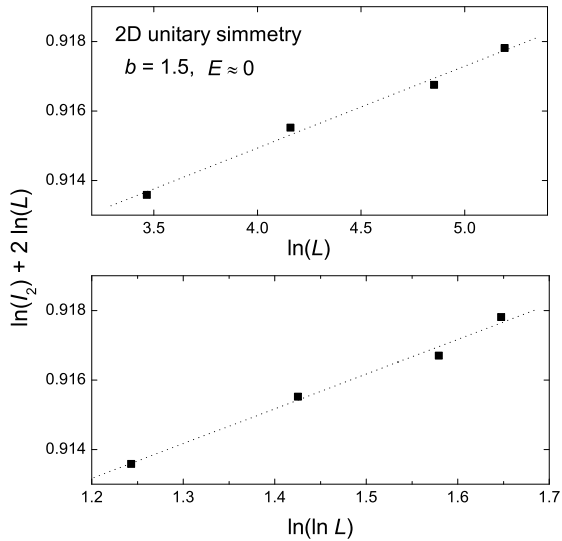}
\caption{Numerical results for the second moments of eigenfunctions. The error bars are smaller than the point size. A fractal law (\ref{standard_scaling}) would imply that for $L\to\infty$ the data points in the top panel would follow a straight line. The law $I_2\propto L^{-2}\ln^{\nu_2} L$ (a possibility suggested by  Eq. (\ref{1 loop}) with $\Pi_{r,r}\approx\ln\ln L$) implies that the points in the bottom panel should follow a straight line.}
\label{figure_weak}
\end{figure}

\paragraph{Weak criticality $(b\gg1)$.} In this limit the model can be mapped onto a non-linear $\sigma$-model with a non-local action \cite{MFD96}:
\be\la{action} S= -\frac{\beta}{16t}\sum_{{\bf r},{\bf r}'} U_{{\bf r},{\bf r}'}\str (Q_{\bf r}Q_{{\bf r}'})+i\omega\sum_{\bf r}\str(Q_{\bf r}\Lambda),\ee
where $Q_{\bf r}$ is an $\bf r$-dependent supermatrix field, $\omega$ is a tunable parameter to be sent to zero in the end of the calculation, and $t=\frac{1}{\pi^2b^2}\ll1$ is the coupling constant. Furthermore, $\Lambda=\mathbf{1}\otimes\diag(1,-1)$ is a matrix of the same size as $Q$, and $\beta$ is a numeric prefactor. If the Hamiltonian $H$ is real (orthogonal symmetry), $\beta=1$ and the size of $Q$ is $8\times 8$; if $H$ is complex (unitary symmetry), $\beta=2$ and $Q$ is $4\times 4$. The non-local kernel of the action is a matrix related to the Hamiltonian variance matrix:
\be\la{U} U_{{\bf r},{\bf r}'}=(a^{-1})_{{\bf r},{\bf r}'} -\delta_{{\bf r},{\bf r}'}(\sum_{{\bf r}_1}a_{{\bf r}_1,{\bf r}_2})^{-1}.\ee
From it we can construct the propagator $\Pi_{{\bf r},{\bf r}'}$ according to
\be\la{Pi} \sum_{\bf r}\Pi_{{\bf r}_1,{\bf r}}U_{{\bf r},{\bf r}_2}=\delta_{{\bf r}_1,{\bf r}_2}-\frac{1}{V},\ee
where $V\sim L^2$ is the volume of the system. In both (\ref{U}) and (\ref{Pi}) the zero-modes are subtracted:
$\sum_{\bf r} U_{{\bf r},{\bf r}'}=\sum_{\bf r}\Pi_{{\bf r},{\bf r}'}=0$.
Assuming a large translationally invariant system, we can work in Fourier space, where Eqs. (\ref{U},\ref{Pi}) mean $\w U(k)=1/\w a(k) -1/\w a(0)$, $\w \Pi(k)=1/\w U(k)$ for $k\neq 0$, and $\w\Pi(0)=0$. For $\alpha=2$,
\be\la{big_integral}
\w a(k)=\int\frac{e^{ikr\cos\phi}}{1+\frac{r^4}{b^4}}rdr\frac{d\phi}{2\pi}.
\ee
To analyze $\w a(k)$ at $bk\ll 1$, we expand $e^{ikr\cos\phi}$ into a power series and integrate it term by term:
\be
\w a(k)=\w a(0) -\frac{k^2}{4}\int\frac{r^3dr}{1+\frac{r^4}{b^4}} +b^2O\Bigl((bk)^4\Bigr).
\ee
The integral over $r$ diverges and hence should be cut off at $r\sim 1/k$. It then yields $(b^4k^2/4)\ln(bk)$, and hence the leading contribution to the propagator is
\be\la{impure}
\w \Pi(k)\approx-\frac{\pi^2}{4}\frac{1}{k^2\ln(bk)}.\ee
This method of computing the integral (\ref{big_integral}) is quite straightforward, if not entirely rigorous. It can be checked, since this integral can also be computed exactly in terms of the Meijer G-function \cite{bateman}. Namely, $\w a(k)=\frac{b^2}{4}G_{3,0}^{0,4} \!\left( \left. \begin{matrix} - \\ 0,\frac{1}{2}, \frac{1}{2},0 \end{matrix} \right| \frac{b^4k^4}{4^4}\right)$; at small arguments, this $G$-function behaves as $G(y^4)=\pi +16(\gamma-1+\log2+\log y)y^2 +O(y^4)$, from where Eq. (\ref{impure}) again follows.

This result is quite different from 1D, where $\w\Pi(k)\propto |k|^{-1}$, i.e. a pure power of $k$, which leads to multifractality. By contrast, Eq.(\ref{impure}) is a power-law modified by a logarithm, which results in the loss of fractal scaling (\ref{standard_scaling}). Indeed, on the one hand, the scaling relation (\ref{standard_scaling}) implies that for small $\Delta_q$
\be I_q\propto L^{-2(q-1)}\Bigl(1-\Delta_q\ln L+\ldots\Bigr).\ee
On the other hand, the perturbative calculation in the $\sigma$-model gives
\be\la{1 loop} I_q\propto L^{-2(q-1)}\Bigl(1-\frac{q(1-q)}{\pi^2 b^2\beta}\Pi_{{\bf r},{\bf r}}+\ldots\Bigr).\ee
The two expressions are consistent if the propagator at coinciding points $$\Pi_{{\bf r},{\bf r}}\approx\frac{1}{(2\pi)^{d/2}}\int^{1/b}_{1/L}\w \Pi({\bf k}) d^d{\bf k}$$ goes as $\ln L$ for large $L$, in which case Eq.(\ref{1 loop}) yields the leading order expression for $\Delta_q$. This is the situation in 1D. In 2D, however, Eq. (\ref{impure}) gives a {\it double logarithmic} leading dependence $\Pi_{{\bf r},{\bf r}}\approx\frac{\pi^2}{4}\ln\ln L$, thus ruling out the possibility of the fractal scaling (\ref{standard_scaling}).  Eq. (\ref{ln_scaling}) with
\be\la{nuq} \nu_q=\frac{q}{4\beta b^2} + o\Bigl(\frac{1}{b^2}\Bigr)\ee
is the most natural law whose perturbative expansion is consistent with the double logarithm found above. In fact, one can extend the perturbative series in Eq. (\ref{1 loop}) to the next order \cite{ROF11}, and find that it is still compatible with Eq. (\ref{ln_scaling}). This means that the logarithmic scaling hypothesis with the above value of $\nu_q$ holds not just in the leading but also in the first subleading order of perturbation theory.

In order to reconcile our result with the apparent power-law scaling found in \cite{C04}, we present in Fig.~\ref{figure_weak} the anomalous part of $I_2$ as  a function of $\ln L$ and of $\ln\ln L$. Although the system size in the simulations was at the limit of currently available computing power, it is still not very large. Hence, we do not expect the lower panel to represent Eq. (\ref{ln_scaling}). Rather, the purpose of this panel is to demonstrate the impossibility of distinguishing between the fractal scaling law and the logarithmic one, based on the available data.

In conclusion, we have demonstrated that although the eigenfunctions of the 2D model with power-law hopping at the metal-insulator transition are neither localized nor delocalized, they are not fractal. In the regime of strong criticality we have found that the scaling of the moments is consistent with a power-law behavior, but the relation between the surface and the corner fractal dimensions rules out a CFT description of the model. In the opposite regime of weak criticality, the anomalous parts of the moments appear to scale as $\ln^{-\nu_q}L$ rather than a power-law, making the standard notion of the wave function multifractality inapplicable. The exact form of the crossover between the two regimes requires further investigation. We believe that our results are relevant for other correlation functions as well as for different 2D models with long-range interactions.

We are grateful to Yan Fyodorov and Vladimir Kravtsov for illuminating discussions. IR and AO acknowledge support from the Engineering and Physical Sciences Research Council [grant number EP/G055769/1]. E.C. thanks the FEDER and the Spanish DGI for financial support through Project No. FIS2010-16430.

\end{document}